\documentstyle[aps,prl,psfig,multicol,epsf,epsfig]{revtex}
\draft
\newcommand{\be}{\begin{equation}}
\newcommand{\ee}{\end{equation}}
\newcommand{\bea}{\begin{eqnarray}}
\newcommand{\eea}{\end{eqnarray}}

\def\ba{\begin{eqnarray}}
\def\ea{\end{eqnarray}}

\begin{document}
\title{Detect Gluon Polarisation at BNL-RHIC Through $J/\psi$ Pair 
Production Process\\[2mm]} 
\author{Jiro Kodaira
\footnote{E-mail: kodaira@theo.phys.sci.hiroshima-u.ac.jp}
\, and \, Cong-Feng Qiao\footnote{
E-mail: qiao@theo.phys.sci.hiroshima-u.ac.jp}}
\address{Department of Physics, Hiroshima University\\
Higashi-Hiroshima 739-8526, Japan}
\maketitle
\begin{abstract}
Our recent study shows that the double spin asymmetry 
of $J/\psi$ pair production is measurable in polarised 
$p-p$ collision at RHIC in near future. And 
hence it enables us to extract the polarised gluon distribution 
function $\Delta G (x)$.

\end{abstract}

\begin{multicols}{2}

Intrigued by the measurement of the European Muon Collaboration
in polarised-target experiment \cite{emc}, an enormous amount of
researches have been carried out on the nucleon spin structure. 
The unpolarised deep inelastic scattering (DIS) experiments
indicate clearly that the gluons share a large portion of the parent 
proton's momentum. However, what portion of proton spin is carried by
gluons is still an open question.
Although there are some efforts \cite{GS,GRVO} to parameterise the polarised
gluon distributions, it is still quite unclear which of these 
parameterisations represents the reality of nature correctly.
To explore this issue, commonly believe that the study of processes other 
than DIS is necessary.
It is now expected that the polarised proton-proton collisions
at BNL relativistic heavy-ion collider(RHIC) 
will provide copious experimental data to unveil the 
polarised parton distributions. 

Up to now, three main kinds of schemes in measuring the gluon polarisation 
are proposed, which are thought to be feasible at RHIC technically. 
Those are
\begin{itemize}
\item{High-$p_T$ Prompt Photon Production}
\item{Jet production}
\item{Heavy Flavor Production}
\end{itemize}
There are advantages and disadvantages in each of these scenarios.
For detailed discussions on this, see recent review paper of 
Ref. \cite{bssv}. 

Quarkonium production and decays have long been taken as an ideal 
means to investigate the nature of QCD and other phenomena. 
Due to the approximately non-relativistic nature, the description 
of heavy quark and antiquark system stands as one of the 
simplest applications of QCD. The very clean signals of quarkonium 
leptonic decays enable the experimental detection with a high precision, 
and therefore quarknoium plays a unique role in investigating 
other phenomena. 

Nevertheless, theoretical description for 
quarkonium production is still premature, although
the heavy quarkonium physics has been investigated for
more than twenty years. In explaining the high-$p_T$ $J/\psi$ 
surplus production discovered by 
CDF group \cite{cdf1,cdf2,cdf3} at the Fermilab Tevatron, 
the color-octet scenario \cite{fleming} was proposed based on 
a novel effective theory, the non-relativistic QCD(NRQCD) \cite{nrqcd}. 
Having achieved the first-step success in explaining the CDF data, 
however, color-octet mechanism also encounters difficulties in 
confronting with other phenomena \cite{rothstein}. 
Due to a recent discovery \cite{qcf1}, the
extent of importance of color-octet contributions in
Charmonium production remains to be unfixed.

During the past years, a series of efforts
have been made on detecting polarised parton distributions
through quarkonium production processes \cite{bs,cp,dr,mty,jk}. 
Most of these investigations are not directly
applicable to RHIC physics, and they are spoiled by
the uncertainties aforementioned.
In recently, it is found \cite{kodaira} that double heavy quarkonium 
production in polarised proton-proton collision provides an 
ideal means to detect the polarised gluon distributions at RHIC, 
and which may at least play a supplemental role to the presently proposed 
program in this end. In literatures \cite{bj,gm}, there were discussions 
of detecting the gluon polarisation by means of double $J/\psi$ production, 
but with main emphasis not on RHIC physics and without presenting the 
analytical expressions for the relative differential cross sections.

The double quarkonium production has
several advantages in reducing the theoretical uncertainties
mentioned above.
(1) By considering double production, the relativistic
corrections and color-octet uncertainties are 
suppressed, especially without or with a lower  
transverse momentum cut \cite{qcf2}. 
(2) The total contribution from the higher excited states 
are also doubly suppressed. 
(3)The higher order QCD correction can be
properly controlled by applying a suitable $p_T$ cut for the 
Charmonium system. 
(4) Since the prevailing partonic process is the gluon-gluon 
fusion into double quarkonium, it stands as a very sensitive 
method in measuring the gluon polarisation. 
In the RHIC energy region, and for 
$p-p$ collision, 
the $q \bar{q}$ initiated process is negligible.
Therefore, one can safely restrict oneself only to the 
gluon-gluon channel as shown in Figure 1. 

\vskip 2mm
\begin{figure}
\begin{center}
\psfig{file=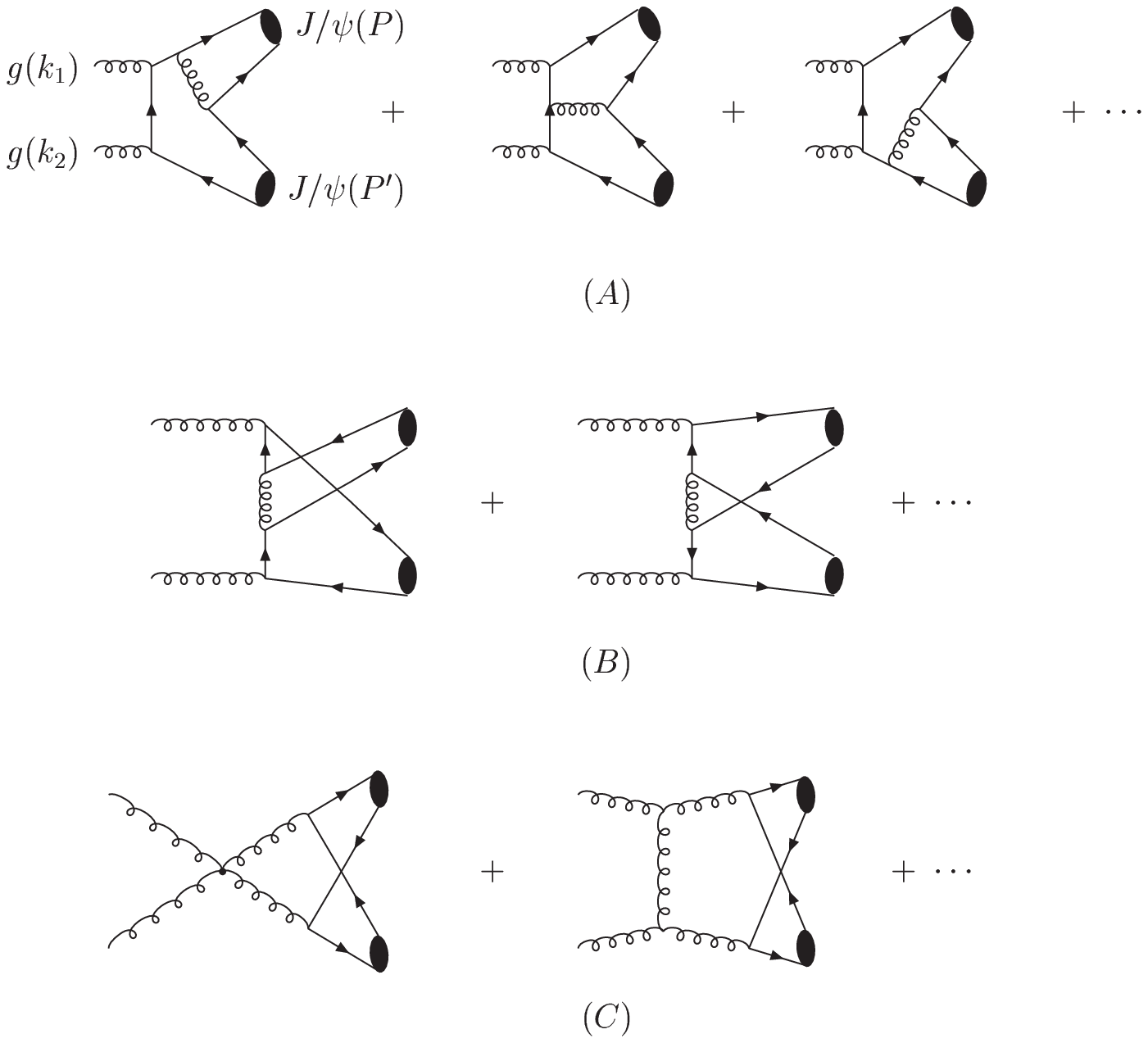, bbllx=70pt,bblly=280pt,bburx=500pt,
bbury=672pt,width=8cm,height=6cm,clip=0}
\end{center}
\caption[]{Typical Feynman diagrams for $g\, + \,g \rightarrow
J/\psi + J/\psi$.}
\end{figure}

When calculating the helicity-dependent matrix elements, one
must project the initial gluons onto definite helicity states.
This can be achieved by taking the ghost free 
expression for the gluon state with helicity $\lambda = \pm$ ,
\bea
\epsilon_{\mu}(k_1,\,\lambda)\;\epsilon^\star_{\nu}(k_1,\,\lambda)
&=&\frac{1}{2}\left[-g_{\mu\nu} + \frac{k_{1\mu} k_{2\nu} + k_{1\nu} k_{2\mu}}
{k_1 \cdot k_2}\right. \nonumber \\
&+& \left. i \,\lambda\;
\epsilon_{\mu\nu\alpha\beta}\frac{k_{1\alpha} k_{2\beta}}
{k_1 \cdot k_2}\right]
\; .
\label{eq0}
\eea
Conventionally, the measurable double spin asymmetry $A$, for $J/\psi$ pair
production, is defined as
\bea
A &=& \frac{d \sigma (p_+ p_+ 
\rightarrow J/\psi J/\psi) - d \sigma (p_+ p_- 
\rightarrow J/\psi J/\psi)}{d \sigma (p_+ p_+ 
\rightarrow J/\psi J/\psi) + d \sigma (p_+ p_- 
\rightarrow J/\psi J/\psi)}\nonumber \\
&=& \frac{E d \Delta\sigma/d^3p}{E d \sigma/d^3p}\; ,
\label{eq1}
\eea
where $p_+$ and $p_-$ denote the helicity projection of
the incident protons being positive and negative, respectively. 
In terms of the gluon densities and the partonic cross sections,
this asymmetry reads
\be A 
     =  \frac{ \int dx_1 dx_2  d \Delta \hat{\sigma}
             \Delta G(x_1, Q^2 ) \Delta G(x_2, Q^2 )}
         {\int dx_1 dx_2  d\hat{\sigma} G(x_1, Q^2 ) G(x_2, Q^2)} \; ,
\ee
where $\Delta G(x, Q^2) = G_+ (x, Q^2 ) - G_- (x, Q^2)$
and $G(x, Q^2) = G_+ (x, Q^2) + G_- (x, Q^2)$
are the polarised and unpolarised gluon distributions defined at the
scale $Q^2$. The unpolarised (polarised) partonic cross section
$\hat{\sigma}$ ($\Delta \hat{\sigma}$) is defined as
\be 
\hat{\sigma} = \frac{1}{4}\ \sum_{\lambda,\lambda'}\ \hat{\sigma}\
(\lambda,\;\lambda'),\;
\Delta\hat{\sigma} = \frac{1}{4}\ \sum_{\lambda,\lambda'}\ 
\lambda\lambda' \ \hat{\sigma} \ (\lambda,\;\lambda').
\ee
The corresponding partonic differential cross sections can be 
found in \cite{kodaira} and \cite{qcf2,hm} in analytical forms.

In doing numerical calculation,
the scale $Q^2$ of the parton distribution function and
the strong coupling constant is taken
to be the transverse momentum of $J/\psi$ for the $p_T$ distributions.
Whereas, in the calculation of the angular 
distribution of the spin asymmetries and the integrated cross sections,
the scale is taken to be $Q^2 = m^2$.
The nonrelativistic relation $m = 2\ m_c$, with
$m_c = 1.5$ GeV, is used and $|R(0)|^2\ = \ 0.8\ \rm{GeV}^3$.

\begin{figure}[tbh]
\begin{center}
\epsfig{file=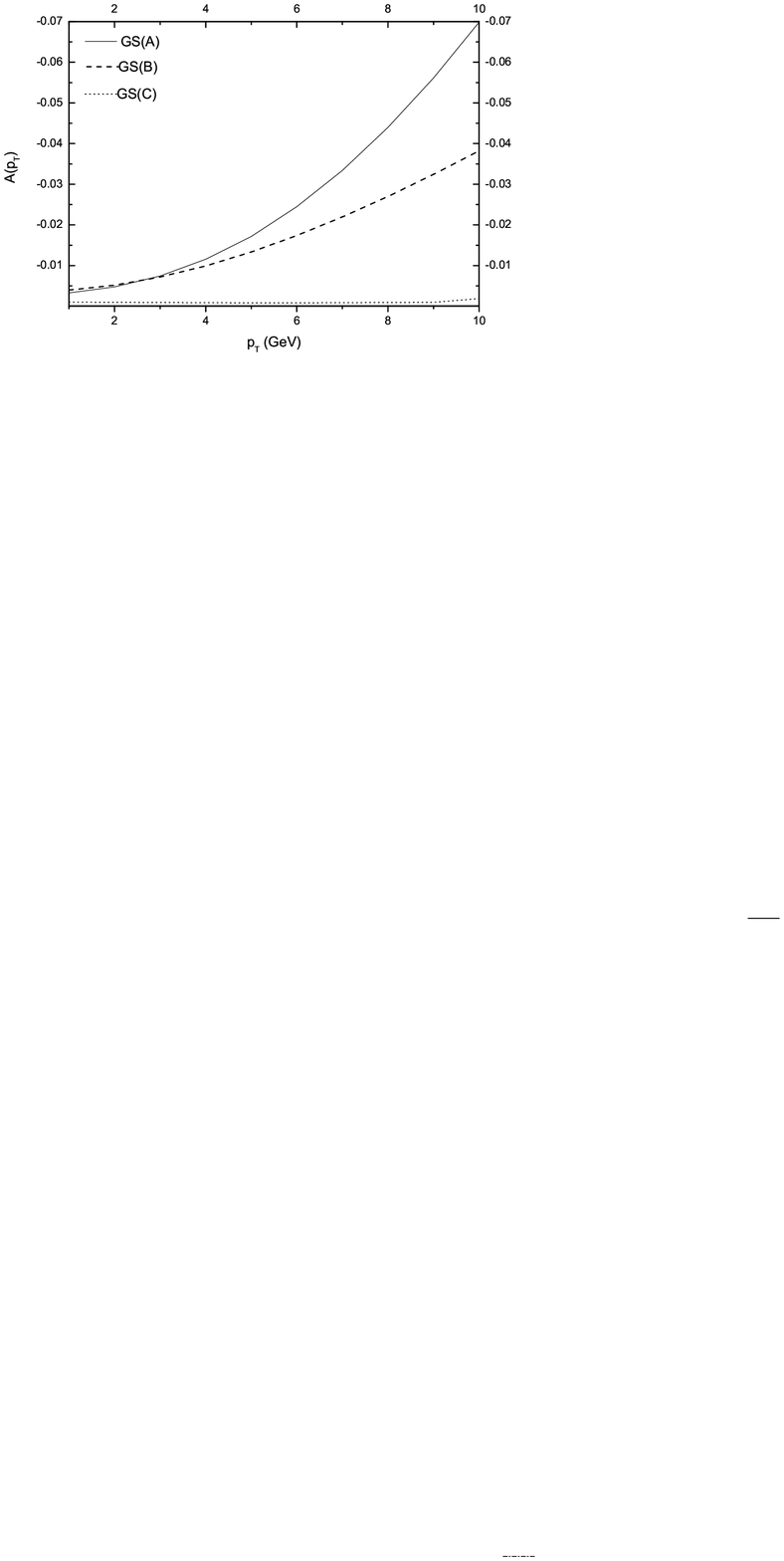,bbllx=30pt,bblly=710pt,bburx=330pt, 
bbury=912pt,width=8cm,height=6cm,clip=0}
\end{center}
\caption[]{Spin asymmetry versus transverse momentum for different 
set of GS polarised parton distributions at colliding energy 
$\sqrt{s} = 500$ GeV.}
\label{graph1}
\end{figure}
In Figure 2 the double spin asymmetry versus the
transverse momentum of $J/\psi$ is presented for different set of 
Gehrmann and Stirling(GS) parameterisations \cite{GS}.
Figure 3 shows the angular distributions of the asymmetry
in the parton center-of-mass frame.
For consistency the MRST parameterisation \cite{mrst} for 
unpolarised gluon distribution is used in drawing 
these figures. One can see that the asymmetries obtained 
by using GS parameterisations
are quite different from each other. 

\begin{figure}[tbh]
\begin{center}
\vskip 1cm
\epsfig{file=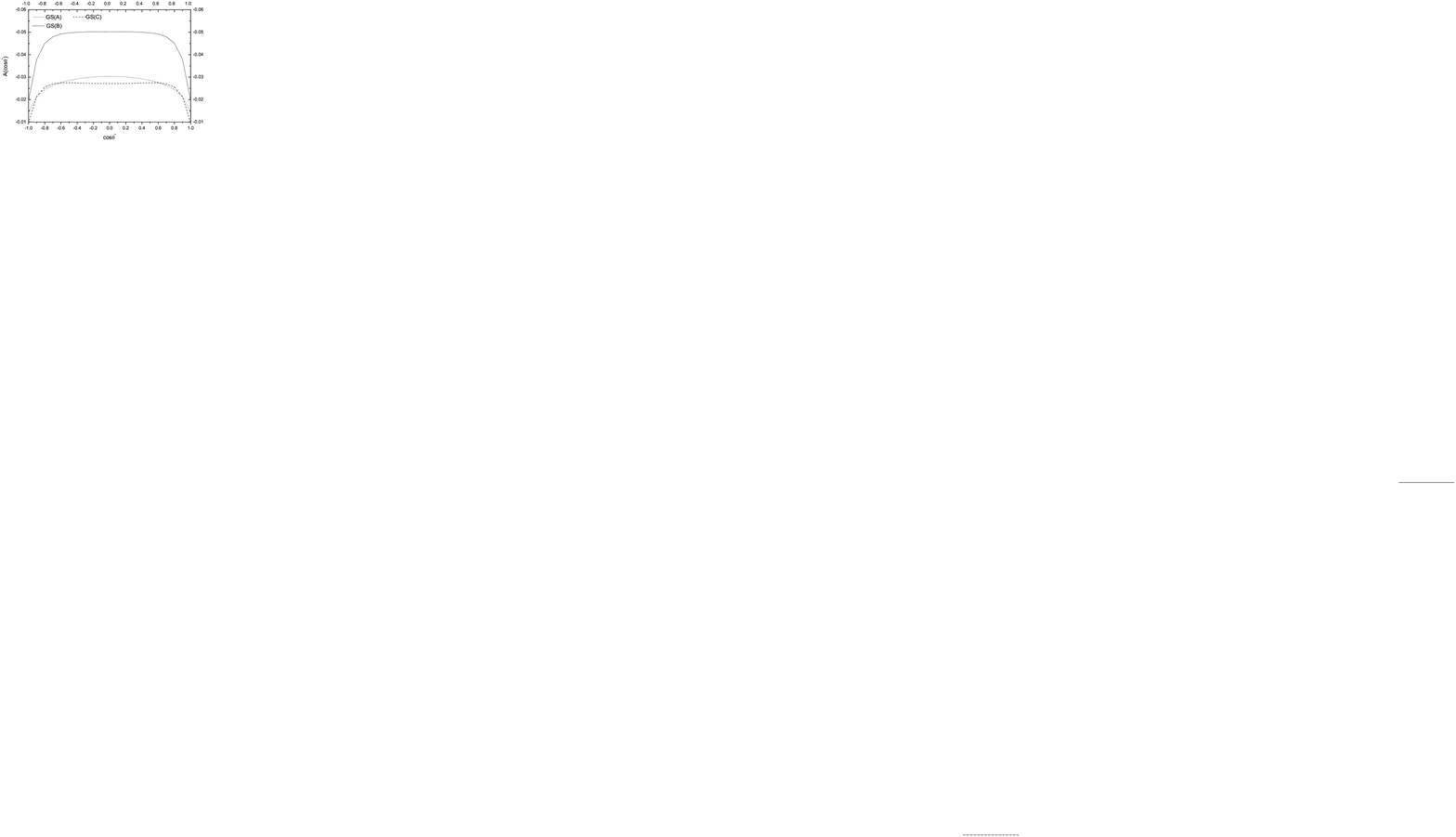,bbllx=20pt,bblly=1024pt,bburx=325pt,bbury=1230pt,
width=8cm,height=6cm,clip=0}
\end{center}
\caption[]{Angular differential asymmetry distribution of the 
$J/\psi$ pair in the parton centre-of mass frame at RHIC with
colliding energy $\sqrt{s}$ = 500 GeV.}
\label{graph2}
\end{figure}    

The total cross sections with different parton parameterisations 
are summarised in table I. The discrepancies among  
these results are not very large as expected. 
\vskip 6mm
{\small{TABLE I. Total cross sections for $J/\psi$ pair production
at RHIC with $\sqrt{s} = 500$ GeV, evaluated with different
parton distributions.}}
\begin{center}
\begin{tabular}{ccc}\hline\hline
\rule[-1.2ex]{0mm}{4ex}&\hspace{-0.6cm} 
$\sigma^{tot}_{\mu^+\mu^-}$ \hspace{0.6cm}&
$\sigma_{\mu^+\mu^-}(|p_T|>1 \;\mbox{GeV})$ \\ \hline
\rule[-1.2ex]{0mm}{4ex} CTEQ5L\cite{cteq} 
\hspace{0.6cm} & $11.8 \; \mbox{pb}$ \hspace{0.6cm}
& $7.3\; \mbox{pb}$ \\ \hline
\rule[-1.2ex]{0mm}{4ex} MRST \cite{mrst}\hspace{0.6cm} & $6.5 
\; \mbox{pb}$\hspace{0.6cm} 
& $4.3\; \mbox{pb}$ \\ \hline
\rule[-1.2ex]{0mm}{4ex} GRV \cite{grv}
\hspace{0.6cm} & $7.4 \; \mbox{pb}$ \hspace{0.6cm}
& $4.7\; \mbox{pb}$ \\ \hline\hline
\end{tabular}
\end{center}
\vskip 5mm

\noindent
Here, the notation $\sigma_{\mu^+\mu^-}$ means that the branching ratio of 
$B(\psi \rightarrow {\mu^+\mu^-}) =  0.0588$, as the practical 
measuring mode to reconstruct the charmonium state, is included.
From the predicted cross sections, we see
that with the integrated luminosity of $800\;\rm{pb}^{-1}$ 
in the future run of RHIC, there will be thousands of $J/\psi$
pair events to be detected, which can certainly give
us some information on the gluon polarisation within the hadron.

In summary, it is found that the $J/\psi$ 
pair production at RHIC may stand as an independent means of
observing the gluon spin distributions within the hadron.
The asymmetry, rather than cross sections, eliminates large amount of
uncertainties which come from the non-perturbative hadronization.
In the future run of RHIC, that is to say with colliding energy of 500 
GeV and accumulated luminosity 800 $\rm{pb}^{-1}$, 
thousands of $J/\psi$ pair events could be detected
and the gluon polarisation could be measured. 
With the expected upgrade of RHIC in future, the concerned 
process would show up to be more important in chasing the 
goal of uncovering the nucleon spin structures. 

The work of J.K. was supported in part by the 
Monbu-kagaku-sho Grant-in-Aid for Scientific Research 
No.C-13640289. The work of C-F.Q. was supported by the 
Grant-in-Aid of JSPS committee.

\end{multicols}

\begin{references}

\bibitem{emc}
European Muon Collab., J. Ashman {\it et al.}, \ Phys. \ Lett. 
{\bf B206}, 364 (1988).

\bibitem{GS} T. Gehrmann and W.J. Stirling, \ Phys. \ Rev. \ D{\bf 53},
\ 6100 (1996).

\bibitem{GRVO} M. Gl\"uck, E. Reya and W. Vogelsang, \ Phys. \ Lett.
\ {\bf B359}, 201 (1995).

\bibitem{bssv} G. Bunce, N. Saito, J. Soffer and W. Vogelsang,
\ Ann. \ Rev. \ Nucl. \ Part. \ Sci.  {\bf 50}, 525 (2000). 

\bibitem{cdf1}
CDF Collaboration, F. Abe {\it et al.}, \ Phys. \ Rev.\ 
Lett. {\bf 69}, 3704 (1992). 

\bibitem{cdf2}
CDF Collaboration, F. Abe {\it et al.}, \ Phys. \ Rev.\ 
Lett. {\bf 79}, 572 (1997).

\bibitem{cdf3}
CDF Collaboration, F. Abe {\it et al.}, \ Phys. \ Rev.\ 
Lett. {\bf 79}, 578 (1997).

\bibitem{fleming} E. Braaten and S. Fleming,
\ Phys. \ Rev.\ Lett. {\bf 74}, 3327 (1995).

\bibitem{nrqcd} G.T. Bodwin, E. Braaten, and G.P. Lepage, 
\ Phys. \ Rev. \ D{\bf 51}, 1125 (1995); {\bf 55}, 
5853(E) (1997).

\bibitem{rothstein} For recent reviews see, for example: 
I. Rothstein, Hep-ph/9911276; M. Kr\"amer,  Hep-ph/0106120.

\bibitem{qcf1} Cong-Feng Qiao, Hep-ph/0202227.

\bibitem{bs} A.V. Batunin and  S.R. Slabospitsky, \ Phys.
\ Lett. {\bf B188}, 269 (1987); 

\bibitem{cp} J.L. Cortes and B. Pire, 
\ Phys. \ Rev.\ D{\bf 38}, 3586 (1988).

\bibitem{dr} M.A. Doncheski and R.W. Robinett, \ Phys.
\ Lett. {\bf B248}, 188 (1990). 

\bibitem{mty} T. Morii, S. Tanaka, and T. Yamanishi, \ Phys.
\ Lett. {\bf B372}, 165 (1996) 

\bibitem{jk} R.L. Jaffe and D. Kharzeev, \ Phys.
\ Lett. {\bf B455}, 306 (1999). 

\bibitem{kodaira} J. Kodaira and C.-F. Qiao, Hep-ph/0207318.

\bibitem{bj} S.P. Baranov and H. Jung, \ Z. \ Phys. 
\ C{\bf 66}, 467 (1995).

\bibitem{gm} T. Gehrmann, \ Phys. \ Rev.\ D{\bf 53}, 5310 (1996).

\bibitem{qcf2} Cong-Feng Qiao, 
\ Phys. \ Rev.\ D{\bf 66} (2002) 057504.

\bibitem{hm} B. Humbert and P. Mery, \ Z. \ Phys. 
\ C{\bf 20}, 83 (1983); \ Phys. \ Lett. {\bf B124}, 256 (1983). 

\bibitem{mrst} A.D. Martin, R.G. Roberts, W.J. Stirling and R.S Thorne, 
Eur. Phys. J. C{\bf 14}, 133 (2000). 

\bibitem{cteq} CTEQ Collaboration, H.L. Lai {\it et al.}, 
Eur. Phys. J. C12, 375 (2000).

\bibitem{grv} M. Gl\"uck, E. Reya and A. Vogt,  \ Z. \ Phys. 
\ C{\bf 67}, 433 (1995).

\end{references}
\end{document}